\documentclass[article,12pt]{revtex4}
\usepackage[applemac]{inputenc}
\usepackage{amssymb}
\usepackage{amsmath}
\usepackage{latexsym}

\def\ie {{i.e.}}

 \def\C{\hspace{3pt}{\rm
l\hspace{-.47em}C}} 
\def\H{{\cal H}} \def\K{{\cal K}}  \def\n{{\cal N}}

\def\uh{\underline{\cal H}}
 \def\uu{\underline{U}}

\def\dg{^{\dag}}

\def\b{\begin{equation}} \def\e{\end{equation}}
\def\bd{\begin{displaystyle}}
\def\ed{\end{displaystyle}}
\def\ba{\begin{array}}
\def\ea{\end{array}}  \def\rw{\rightarrow}
\def\bee{\begin{enumerate}} \def\eee{\end{enumerate}}
\def\ees{\end{eqnarray*}} \def\be{\begin{eqnarray}}
\def\ee{\end{eqnarray}} 
\def\le{\langle}
\def\re{\rangle}

\newcommand{\ket}[1]{\mid\!#1\,\rangle}
\newcommand{\bra}[1]{\langle\,#1\!\mid}

\newcommand{\Id}[1]{\mathrm{Id}_{#1}}

\newcommand\hs[1]{\hskip#1pt} \def\bes{\begin{eqnarray*}}
\newcommand{\lbl}[1]{\label{eq: #1}} \newcommand{\rf}[1]{\ref{eq: #1}}

\begin{document}

\title{Krein space quantization in curved and flat spacetimes
\footnote{Preprint to be published in Journal 
of Physics A (http://www.iop.org). }}

\author{
T. Garidi$^{1,2}$, E. Huguet$^{2,3}$, J. Renaud$^{1,2}$}
\affiliation{$1$ - LPTMC, Universit\'e Paris 7-Denis Diderot, boite 7020
F-75251 Paris Cedex 05, France.\\
$2$ - F\'ed\'eration de recherche APC, Universit\'e Paris 7-Denis Diderot, 
boite 7020, F-75251 Paris Cedex 05, France.\\
$3$ - GEPI, Observatoire de Paris, 5 place J. Janssen, 92195 Meudon
Cedex, France.} \email{garidi@kalymnos.unige.ch, 
eric.huguet@obspm.fr,  renaud@ccr.jussieu.fr.
}

\date{\today}

\begin{abstract}
We reexamine in detail a canonical quantization method \emph{\`a la}
Gupta-Bleuler in which the Fock space is built over a so-called Krein
space. This method has already been successfully applied to the
massless minimally coupled scalar field in de Sitter spacetime for
which it preserves covariance. Here, it is formulated in a more
general context. An interesting feature of the theory is that, 
although the field is obtained by
canonical quantization, it is independent of  Bogoliubov transformations.
Moreover no infinite term appears in the computation of $T^{\mu\nu}$ 
mean values
and the vacuum energy of the free field vanishes: $\bra{0} T^{00}
\ket{0}=0$. We also investigate  the behaviour of the Krein
quantization in Minkowski space for a theory with interaction. We show
that one can recover the usual theory with the exception that the
vacuum energy of the free theory is zero.
\end{abstract}


\maketitle

\section{INTRODUCTION}
Recently a covariant quantization of the massless minimally coupled 
scalar field on  de Sitter (dS) spacetime has been carried out \cite{grt}. 
Indeed, the quantization scheme used is not restricted 
to the dS spacetime, in this paper we describe it in a more general 
context.

Although quantum field theory (QFT) on a curved classical 
background
has already been extensively studied \cite{birreldavis,wald}, 
various problems are still open.  This is even true for 
the de Sitter
space, which is of current importance. The dS
space really appears as one of the simplest curved spacetimes:
it is
maximally symmetric and 
offers the 
opportunity  of controlling the
transition to the flat spacetime by the so-called contraction 
procedure \cite{garhugren}. In view
of these properties  dS spacetime should at least be considered as
an excellent laboratory. Nevertheless, even for this very simple case, one
encounters difficulties in defining quantum fields, including the free fields.

First, the one-particle state space is usually taken as the
carrier space of an unitary irreducible representation of the symmetry
group (elementary system in Wigner's sense). In the massive Minkowskian case,
the carrier space is a Hilbert space which can be selected on physical
grounds by the  condition of energy positiveness. But since the energy 
concept
is defined only for stationary spacetimes, it turns out that, in
general, one cannot characterize in a unique way a preferred invariant
subspace of solutions of the field equation. This kind of difficulty
actually corresponds to the choice of a vacuum state among a family 
of the so-called $\alpha$-vacua introduced in 
\cite{allen1, chernikov, mottola}. The choice of a vacuum among
 the family of  
$\alpha$-vacua  has been recently largely
discussed in various context related to inflation 
\cite{daniel1,daniel2},  
dS/CFT correspondence \cite{antoniadis,mazur,bousso,tolley}, or QFT
\cite{banksmannelli,einhornlarsen,goldsteinlowe,garhugren} and no
conclusive arguments which would singled out a preferred vacuum state
have been given so far.

But it comes worse. A famous example is given by Allen's no-go theorem
concerning the so-called massless minimally coupled scalar field in dS 
space
for which it has been claimed that no invariant vacuum
exists. That is to say that no covariant Hilbert space quantization
is possible \cite{allen1}.  It actually happens that no invariant
Hilbert space $\H_p$ containing every regular solutions of the
field equation, with initial conditions having a compact support, can
be defined at all.  This difficulty is still present for the dS massless
spin-2 field.

In this paper, we present a new version of the canonical quantization
which allows covariant quantization even in situations where the usual
method fails, including the dS massless minimally coupled field and
spin-2 massless field. This construction is of the Gupta-Bleuler type
and the set of states is different from the set of physical states.
Instead of having a multiplicity of vacua, we have several
possibilities for the space of physical states and only one field and
one vacuum 
the latter being invariant under Bogoliubov
transformations.  So the usual ambiguity about vacua is not suppressed but displaced.
Our construction gives a framework (the Krein space) where all  objects such  as field, Bogoliubov transformations and vacuum can be treated together in a very straightforward way (for an alternative approach, see \cite{higu1,higu2,marolf}).
Moreover this field presents an interesting property
linked to the cosmological constant problem: the vacuum energy of the free field
vanishes without any reordering nor regularization.

We also investigate how this method behaves in the
Minkowski space when interaction is present. In fact, 
a successful quantization scheme in the dS space must 
lead to the common results when applied in the 
usual flat case.

Sec. II gives a brief overview of the problem of quantizing a free
scalar field in a fixed curved spacetime background. In Sec. III, we
introduce a simple and basis independent rigorous formulation in which 
the minimal requirements for a canonical quantization are formulated
We explain in Sec. IV a new method  based on weaker conditions, 
which do
not prohibit  negative norm states in the definition of the field.
Actually the quantization will be done in the larger framework of
a Krein space which will  preserves the covariance.  The Sec. V is 
devoted to the flat case with 
a self-coupling.

\section{FREE-FIELDS IN A CURVED SPACETIME}

Let us  consider a spacetime $M$, that is a
pseudo-Riemannian manifold endowed with a metric $g_{\mu \nu}(x)$
with a signature $(+, -, -, -)$. We  restrict ourselves to the free
scalar field whose Lagrangian density reads
$${\cal L}=
g^{\mu\nu}\partial_\mu\phi\partial_\nu\phi-\zeta^2\phi,$$ where
$\zeta$ is a real coupling parameter whose meaning will not be
discussed here. The corresponding field equation reads
\begin{equation} \label{equation}
\Box \phi + \zeta^2\phi = 0\,.
\end{equation}

The spacetime is supposed to be globally hyperbolic \cite{fulling,isham}.
Thus, the field equation admits Cauchy space-like surfaces
$\Sigma$ and a two-point function $\tilde G$ which satisfies
\begin{equation}
\label{2pt}
\phi(x)=\int_\Sigma\tilde
G(x,x')\stackrel{\leftrightarrow}{\partial_\mu}\phi(x')\,d\sigma^\mu(x'),
\end{equation}
for each regular solution $\phi$. Although
it is a purely geometrical object,  the function $\tilde G$ is
referred to  as the commutator of the field. This function $\tilde
G$ is a solution of the field equation in each of its variables
and, for a given $x$, its support in $x'$ belongs to the causal
cone of $x$. Hence, the classical field $\phi$ is determined by
its initial conditions on the surface $\Sigma$.  Consistently 
with (\ref{2pt}) the space of
solutions of the field equation is endowed with the Klein-Gordon
scalar product,
\begin{equation}\label{pskg}
\le\phi_1,\phi_2\re=i\int_\Sigma\phi^*_1
\stackrel{\leftrightarrow}{\partial_\mu}\phi_2\,d\sigma^\mu,
\end{equation}
which is independent of the Cauchy surface $\Sigma$. Note that (\ref{pskg})
may be indefinite. In flat space
time, this scalar product is nothing but the natural scalar product
for the functions defined on the mass hyperboloid in the momentum
space.

The usual (L$^2$) scalar product 
\begin{equation}\lbl{psldeux}
(f_1,f_2)=\int_Mf_1^*(x)f_2(x)d\lambda(x),
\end{equation}
where $d\lambda$ is the natural measure
$d\lambda(x)=\sqrt{|g(x)|}d^4x$, is also defined for functions on the 
manifold. 
Contrary to the Klein-Gordon
product, this product is positive definite and the functions $f_1$
and $f_2$ are not necessarily solutions of the field equation.

Some very important spacetimes, Minkowski of course
but also  de Sitter and anti de Sitter spacetimes, admit  a
large isometry group $G$. This group acts in a natural way on the
solutions through $U_g\phi(x)=\phi(g^{-1}x)$ for $g\in G$. The
Klein-Gordon and the L$^2$ products are  left invariant under this
action, thus one has
$$\le U_g\phi_1,U_g\phi_2\re=\le\phi_1,\phi_2\re,$$and$$
(U_gf_1,U_gf_2)=(f_1,f_2).
$$

Quantizing such a field amounts to build an operator-valued
distribution $\varphi$, rigorously defined by $\varphi(f)$, where $f$
is a test function i.e., a ${\cal C}^\infty$ function with compact
support on $M$.  For convenience, we will use the common abusive
notation $\varphi(x)$. The field must satisfy minimal requirements to
be acceptable: it must verify the field equation, commute for space-like
separations (micro-causality) and satisfy the
covariance condition
$$\uu_g^{-1}\varphi(x)\uu_g=\varphi(g\cdot x),\hs{10}\forall x\in
M,\;\forall g\in G.$$

It is well known that the quantization procedure based on the Hilbert
space structure and satisfying the above conditions already fails
in many situations. For instance, in QED, the only way to preserve
(manifest) covariance and gauge invariance in canonical quantization
is to use the Gupta-Bleuler method.  Similarly, it has been shown
\cite{allen1} that, for the minimally coupled scalar field ($\zeta =
0$ in (\ref{equation})) in de~Sitter spacetime, no Hilbert space can
be defined for the quantization. The price to pay in both cases in
building a covariant quantum field is the appearance of unphysical
negative norm states.  It is commonly accepted that the very reason
for that impossibility relies, in the case of electrodynamics, upon
the invariance of the Lagrangian under a gauge transformation.
Actually this also occurs for the dS massless minimally coupled scalar
field, where the transformation ``$\phi\to\phi + \mbox{constant}$'',
viewed as a gauge transformation, is a symmetry of the Lagrangian.

In this paper we present, a method which, within the framework of
canonical quantization, generalizes the Gupta-Bleuler procedure to
situations in which the occurrence of  negative norm states is
unavoidable in order
to achieve a covariant quantization. More precisely, as for the
Gupta-Bleuler method, the field will be written in terms of a
Krein space  and  the task will consist in finding the
conditions which define the one-particle sector.

\section{CANONICAL QUANTIZATION}

Before presenting the Krein space method, let us indicate
a rigorous and intrinsic formulation of the usual canonical quantization.
 This material will be used
for the Krein space quantization as well. Actually, we will see that
our method is also a canonical quantization which only differs from
the usual one by the representation of the canonical commutation
relations.

We first give an 
heuristic overview of the canonical quantization method, which
can be summarized as follows.
Let $\phi_k$ be a family of solutions (supposed to be known) of
the field equation and verifying
\begin{align}\label{caqu1}
 \le\phi_k,\phi_{k'}\re=&\, \delta_{kk'},\\
\label{caqu2} \le\phi_k,\phi^*_{k'}\re=&\, 0,\\
\label{caqu3} \sum_k\phi_k(x)\phi^*_k(x')-\phi^*_k(x)\phi_k(x')=&\, -i\tilde
G(x,x').
\end{align}
The choice of the modes ($\phi_k$) determines the space $\H_p$
they generate as an Hilbertian basis. That space is usually called
the one-particle sector, although the word ``particle'' may be
confusing. The question of how to make this  choice is answered
for stationary spacetimes (Minkowski) by the fact that one can
find a Killing vector field $X$ which is time-like at each point,
and thus naturally  defines  a Hamiltonian for the quantization
space: $iX$. The space $\H_{p}$ is then chosen in such a  way
that the Hamiltonian admits a positive spectrum; the modes are
generally chosen as eigenvectors, with positive eigenvalues, of
that operator. The corresponding modes $\phi_k$ are said to be
positive frequency modes.

Whenever the considered spacetimes are non stationary, one is
left with the orthonormality conditions alone to characterize
$\H_{p}$. Unfortunately, these do not uniquely determine the one-particle 
space. Actually, if the $\phi_k$'s satisfy the above
conditions so do the family $\tilde \phi_k=A\phi_k-B\phi^*_k$ with
$|A|^2-|B|^2=1$ (Bogoliubov transformation). The question of which
family should be preferred is not trivial since the two-point
functions, (thus also the expectation values) will depend on that
choice. Because the vacuum is defined through the modes, it  has
become usual to say that the choice of the modes is equivalent to
the choice of the vacuum.  We will see that our point of view is
different.

The third condition (\ref{caqu3}) ensures the completeness of the family
$\{\phi_k\}$. The space spanned by the $\phi_k$'s and the
$\phi_k^*$'s contains all the regular solutions of the field
equation for which the initial conditions have a compact support.

 The quantum field is now defined through
\begin{equation}\label{quantumfield}
\varphi(x)=\sum_k\phi_k(x)A_k+\phi_k^*(x)A\dg_k,
\end{equation}
where $A_k$ and $A_k\dg$ are operators verifying the canonical
commutation relations (ccr)
$$[A_k,A\dg_{k'}]=\delta_{kk'},\hs{10}
[A_k,A_{k'}]=0,\hs{10}[A\dg_k,A\dg_{k'}]=0.  $$ Our notation
departs from the usual one (with small letters) because the above
relations are the starting point of both, the usual and what we
shall call in the sequel the Krein canonical quantization. These
two quantizations will differ only through the representation of
these relations, that is through the realization of the $A_k$ as
operators acting on some vector space. Before any representation
has been set, it is clear that, at least formally, such a field
satisfies the field equation and the causality requirement since
the ccr imply
$$[\varphi(x),\varphi(x')]=-i\tilde G(x,x').  $$
The covariance condition seems not so clear but we shall use a
formalism  which makes it obvious, the only requirement being
that the space $\H_{p}$ is closed under the action of the isometry
group. Recall that the space $\H_{p}$ is defined through the 
square-summability
requirement
$$\H_{p}=\left\{\sum_{k}c_{k}\phi_{k};\;
\sum_{k}|c_{k}|^{2}<\infty\right\}.
$$

Regarding the usual canonical quantization, the representation of the
ccr is realized in the Fock space $\underline{{\H}_p}$ (see the
appendix) by $$A_k=a_k:=a(\phi_k)\mbox{ and }
A_k\dg=a_k\dg:=a\dg(\phi_k).  $$ It follows from (\ref{quantumfield})
and the linearity and anti-linearity of the creation and annihilation
operators respectively that
\begin{equation*}
\varphi(x) = a(p(x)) + a\dg(p(x)),
\end{equation*}
where $p(x)$ is  defined through
$$p(x)=\sum_k\phi_k^*(x)\phi_k.
$$

The key point is that $p$ can be defined without reference to the
basis. For that remind that $p$ is a ${\H}_p$-valued
distribution. Consequently, it must be smeared by test functions
(which are chosen to be real) $f\in{\cal C}^\infty_o(M)$. The
rigorous definition of $\varphi$ reads \bes \varphi(f)&=& \int
f(x)
\varphi(x) d\mu(x)\\
&=&\sum_{k}(\phi_k^{*},f)a_k+\sum_{k}(\phi_k,f)a^{\dag}_k,
\ees
where the parenthesis stands for the L$^2$ product. Using the linearity
one obtains
\begin{equation*}
\varphi(f) = a\left(\sum_{k}(\phi_k,f)\phi_k\right) +
a^{\dag}\left(\sum_{k}(\phi_k,f)\phi_k\right).
\end{equation*}
Defining
\begin{equation}\lbl{proj}
p(f)=\sum_{k}(\phi_k,f)\phi_k \in\H_p,
\end{equation}
one finally  gets
\begin{equation}
\varphi(f)= a\left( p(f)\right)+a^{\dag}\left(
p(f)\right),\lbl{smfield}
\end{equation}
where $p$ is a  $\H_p$-valued distribution. A base free definition
of $p$  is that   $p(f)$ is the only element of the Hilbert space
$\H_p$ for which
\begin{equation} \le
p(f),\psi\re=(f,\psi),\hs{20}\forall\psi\in \H_{p},\lbl{tilde}
\end{equation}
where the scalar products  $\le\ ,\ \re$ and $(\ ,\ )$ are
respectively defined in Eq. (\ref{pskg}) and  Eq. (\rf{psldeux}).
The expressions (\rf{smfield}) and (\rf{tilde}) then allow to give
a rigorous definition of  the field which depends on the choice of
$\H_p$ but not on the basis given by the modes. The covariance of
the distribution $p$ easily follows from this definition under the
sole   hypothesis that $\H_p$ is closed with respect to the action
of the group $G$. More precisely, we show that $p$ intertwines the
natural representations $U$ of $G$ over  ${\cal C}^\infty_o(M)$
and $\H_p$~:
\begin{equation}\lbl{picov}
U_gp=p U_g\,.
\end{equation}
Indeed, for $\psi\in\H_p$ one has \bes
\le U_gp(f),\psi\re&=&\le p(f),U_g^{-1}\psi\re\\
&=&(f,U_g^{-1}\psi)\\ &=&(U_g f,\psi)\\ &=&\le p(U_gf),\psi\re.
\ees The representation $U$ is extended to a representation  $\uu$
on $\underline{{\cal H}_p}$, and therefore \bes
\uu_g\varphi(f)\uu_g^{-1}&=&a\left(U_gp(f)\right)+
a^{\dag}\left(U_gp(f)\right)\\ &=&a\left(p(U_gf)\right)+
a^{\dag}\left(p(U_gf)\right)\\ &=&\varphi(U_gf). \ees {\it Thus,
the usual canonical quantization, and its covariance, rests on the
existence of a space of solution $\H_p$ in which the scalar
product is positive, which is closed under the group action and
such that $\H_{p}+\H_{p}^{*}$ contains all the regular solution of
the equation}. It is important to mention a technical point: in
definition  (\rf{tilde}) it is assumed that the map
$\psi\mapsto(f,\psi)$ is continuous from $\H_p\rw\C$. This
condition is not always realized  and must be checked
individually.

Within the canonical quantization there are  infinite terms which
appear in the energy-momentum tensor and its expectation values
for some field states, especially in the vacuum. This calculation
rests on the ccr and on the definition of the vacuum
$a_k|0\re=0$. In the following section,  we  describe a new
quantization verifying the ccr but with a different vacuum:
we will show that these infinite terms actually disappear.

\section{KREIN SPACE QUANTIZATION}

This quantization is of the Gupta-Bleuler type in the sense that
one will distinguish the space on which the observables are defined
(called the total space) from the subspace of physical states
on which the average values of the observables are computed. Note
that the total space  is equipped with an indefinite inner product
which means that some (unphysical) states have a negative norm.

The original Gupta-Bleuler quantization was invented  in order to
preserve the (manifest) covariance in presence of gauge invariance. Not
surprisingly, a similar method works in the case of the dS massless 
minimally
coupled scalar field  (contrary to the usual
canonical quantization) in which   gauge invariance also
occurs.

We use a set of modes satisfying the conditions  given above.
The main difference is that we no longer require  that  the space
$\H_p$ is closed under the group action. It is enough that
the larger space  $\H=\H_p+\H_p^*$  is closed under the latter, 
which is obviously
 a weaker condition. Again, the field is defined by
Eq. (\ref{quantumfield}). Now,  the representation
of the ccr is obtained in the following way. We consider
the Fock space  $\uh$ \cite{mintchev} constructed on the Krein space
\cite{bognar} $\H=\H_p+\H_p^*$, where $\H_p^*$ is the anti-Hilbertian
space generated by the set ${\phi_k^*}$. Let us set
$a_k=a(\phi_k)$ and $b_k=a(\phi_k^*)$ then
$$A_k=\frac{1}{\sqrt2}(a_k-b\dg_k) \mbox{ and
}A_k\dg=\frac{1}{\sqrt2}(a_k\dg-b_k).$$

Since these operators verify the ccr, we obtain a further
realization of these relations  which yields a new definition of
the field. A significant difference with the usual representation
is that the operators $A_k$ do not verify $A_k|0\re=0$, where  $|0\re$ is the 
Fock vacuum.
 We can now
point by point  summarize the discussion of the above paragraph. The
first two conditions (field equation and micro-causality) are
still verified since they do not depend on the involved
representation of the ccr.

In order to show the  covariance, we start by rewriting the field
in the smeared form as in the previous section. We define the
distribution $p$  taking values in  $\H$ in the following way. For
any function $f$, $p(f)$ is the unique element of $\H$ such that
\begin{equation}
\lbl{defp}\le p(f),\psi\re=(f,\psi)\ \ \forall\psi\,\in\,
\H.\end{equation}
The existence of $p(f)$  is subject to a technical
condition on $\H$: the continuity for each $f$ of the map $\psi\mapsto
(f,\psi)$, $\H\to\C$. 
Formally, $p(f)=\int p(x)f(x)dx$, yields the formal definition
$\le p(x),\psi\re=\psi(x)$. It is easy to verify that the kernel
$\tilde G$ of  $p$    is given by:
$$\le p(x),p(x')\re=-i\tilde G(x,x').$$
Similarly, one can easily  check that   the
field reads
\begin{equation}\lbl{champ}\varphi(x)=a(p(x))+a\dg(p(x)).
\end{equation}
The advantage of the formulation (\rf{champ}) is that the
covariance of the field is shown as above (Eq.
(\rf{picov}) and the corresponding demonstration), the only
condition being that the space $\H$ (not $\H_p$) must be closed
under the group action. This is a remarkable advantage of the
method since, as discussed in details in 
\cite{dbr2,dbr3,grt}, situations do arise in which the space $\H$
does not admit a closed invariant Hilbert subspace, and consequently
it is not possible to define $\H_p$ in a covariant way.

The total space $\H$ contains negative norm states and cannot be
considered as the  space of physical states. The latter must be a
subspace $\K$ which is not necessarily a Hilbert space but a closed
invariant subspace. On $\K$, the inner product is positive,
possibly indefinite. In this case the space $\n=\K\cap\K^{\perp}$
is the set of unobservable gauge states and completes the
so-called Gupta-Bleuler triplet
$$\n\subset\K\subset\H\,.$$
The space $\K/\n$ is the set of physical states stricto sensu.
This yields a condition for a self-adjoint operator to be an
observable: gauge change must not be observed (see \cite{dbr2}).
 Moreover, when $\n$
is non trivial, the representation of the group is not
irreducible, but only indecomposable. That is the space $\K$ is an
invariant subspace of $\H$ but it is not invariantly complemented
in $\H$: there is no invariant subspace $\cal L$ such that
$$\H=\K\oplus\cal L.$$
The same  scheme occurs for $\n$ in $\K$. In the usual case 
(\emph{e.g.} massive scalar minkowskian field) the physical 
space $\K=\H_{p}$ is  invariantly complemented since
$$\H=\H_{p}\oplus\H_{p}^{*}.$$

Note that this field is, by construction (it depends only on $\cal H$), independent of any
Bogoliubov transformations which only modify the set of physical states.
The physical states space  depends on the considered  space time, 
but it also depends on the observer: the physical states of an accelerated observer in Minkowski space are
different from those of an inertial observer (Unruh effect). In our formalism, the same representation of the field
can be used for both cases. Even more, following Wald, there is a case where
the Krein space appears in a very natural way. ``For a space-time which is asymptotically stationary in both the past and the future we have two natural choices of vacua, and the $S$ matrix should be a unitary operator between both structures" \cite{wald} p.66. As pointed out by the author this is not possible if the two vacua are not equivalent. Once again our construction furnishes a stage where all these objects can be considered together.  The difference with the algebraic approach is that states (physical or not) are naturally realised as wave functions. This is not the case in the algebraic theory where this representation requires a choice wich is equivalent to the choice of vacuum.

This method  not only allows to quantize some fields resisting to
the usual procedures, but it presents a very interesting property:
the vacuum energy vanishes without any reordering and no infinite
term appears in the calculation of mean values of $T_{\mu\nu}$. 
Moreover, this is independent of the curvature.
The point is
$$[b_k,b\dg_k]=[a(\phi_{k}^{*}),a\dg(\phi_{k}^{*})]=
\le\phi_{k}^{*},\phi_{k}^{*}\re=-1.
$$
The vacuum is now the Fock vacuum and a straightforward computation leads
to
$$
\le0|(A_kA_k\dg+ A\dg_kA_k)|0\re=0,
$$ 
from which the result
follows easily (see \cite{grt} for details). This shows that this type of 
quantization does not
require any  regularization of the $T_{\mu\nu}$'s: the average
value of the stress tensor in the vacuum state are not only
finite, they actually are zero.

Thus, under weaker condition than in the usual canonical scheme, the
Krein space quantization allows to define a causal and covariant field
which features a remarkable automatic regularization mechanism. It has
been successfully applied to the case of the massless minimally
coupled scalar field in de Sitter space (with field equation $\Box
\phi=0$)\cite{grt}. This case is very important because it also
appears in the dS massless spin-$2$ case, that is the graviton on de
Sitter space.

The Krein space quantization  gives rise to unphysical states.
An essential question  then arises: how to
select the physical states ?  

First of all, it should be emphasized that in presence of gauge
invariance, the occurrence of Krein structure is unavoidable. Even the
standard Gupta-Bleuler quantization involves a Krein space.  The
one-particle physical state space must be a positive inner
product subspace of $\H$, closed under the group action. Generally,
this choice is not unique and we end up with the usual ambiguities
which plague quantum fields theories in curved spacetimes. Note
however the different standpoints: the vacuum state is here considered
as the vacuum state of the Fock space. It is invariant under
Bogoliubov transformations which are merely a simple change of the
physical state space. It must be emphasized that, in presence of gauge
invariance, that space would not be a Hilbert space.

\section{AN INTERACTING THEORY IN MINKOWSKI SPACETIME}\label{interac}

As indicated in the previous sections, our quantization scheme has
already been applied to the dS minimally coupled massless scalar
field. In that context it allows to quantize the free field without loosing
the covariance. Interesting features are that the vacuum is invariant
under a Bogoliubov transformation and no infinite term appears in the
computation of $\bra{0} T^{00} \ket{0}$.

However, only free fields have been studied so far. This yields the 
natural question: how such features behave when interaction is
taken into account.

More precisely let us investigate the way in which the Krein
quantization method can be extended to the simple example of an
interacting scalar field (called k-scalar field in the sequel) in
Minkowski spacetime defined through the Lagrangian density
\begin{equation*}
{\cal L}=
g^{\mu\nu}\partial_\mu\phi\partial_\nu\phi-m^2\phi-V(\phi).
\end{equation*}
We will see in the following that, with a standard choice for
$V(\phi)$, we recover the usual results of interacting field theory.

The unitarity condition, which we discuss now, takes a different form
in our Krein space context. Here, in absence of gauge invariance, we
have $\K=\H_ p$. The free field Fock space is denoted by
$$\underline{{\cal H}} = \underline{{\cal H}_p \oplus {\cal H}_p}^*$$
Moreover the space of physical states $\underline{{\cal H}_p}$ is
closed and positive, hence complementarisable $$\underline{{\cal H}}=
\underline{{\cal H}_p} \oplus \left(\underline{{\cal H}_p}\right)^\perp\, .
$$
 The positive energy (resp. negative energy) one-particle
eigenstates of the free Hamiltonian are denoted as~$\ket{k}$ (resp.
$\ket{\bar k}$). The corresponding symetrized and normalized
many-particle states are denoted by $\ket{k_1,\ldots,k_m;\bar
k_1,\ldots,\bar k_n}$. We shall adopt the abbreviated notations
$$\ket{\alpha_+} \equiv \ket{k_1,\ldots,k_m},$$
$$\ket{\alpha_-^{(n)}}
\equiv \ket{k_1,\ldots,k_m;\bar k_1,\ldots,\bar k_n}, \,\mbox{n} \ne 0.$$
A state without sub- or superscripts has an arbitrary number of
$k$ and $\bar k$.

If one proceeds to the computation of the $S$-matrix as usual:
\begin{equation}\label{st} S=:\exp\left(i\int V(\phi)\right):\ ,\end{equation}
one obtains obviously an operator $S$ which is Krein-unitary.
In view of the non-positiveness  of the scalar product, the unitarity of
the $S$ matrix over a Krein space does not lead to the usual relations
\begin{eqnarray}
\sum_{\{\beta\}} \left |\bra{\beta} S \ket{\alpha}
\right |^2  &=& 1 ,\label{unita1} \\
\sum_{\{\beta\}} \left |\bra{\alpha} S \ket{\beta}
\right |^2  &=& 1. \label{unita2}
\end{eqnarray}
 Indeed,
for $S$ such that $S^\dagger S= S S^\dagger  = \Id{}$ (Krein-unitarity) 
one has
\begin{equation}\label{kunita1}
\sum_{\{\alpha_+\}} \left |\bra{\alpha_+} S \ket{\alpha}
\right |^2  +
\sum_{\{\alpha_-^{(n)}\}} (-1)^n \left |
\bra{\alpha_-^{(n)}} S
\ket{\alpha}
\right |^2 = 1
\end{equation}
\begin{equation}\label{kunita2}
\sum_{\{\alpha_+\}} \left |\bra{\alpha} S \ket{\alpha_+}
\right |^2 +
\sum_{\{\alpha_-^{(n)}\}} (-1)^n \left |\bra{\alpha} S
\ket{\alpha_-^{(n)}}
\right |^2 =  1 .
\end{equation}

Obviously, Eqs.(\ref{kunita1}, \ref{kunita2}) cannot share the same
usual probabilistic interpretation as Eqs. (\ref{unita1}, \ref{unita2})
and thus the Krein-unitarity has no physical meaning. If one
ignores the problem and compute, for $V(\phi)=\lambda\phi^4$ for
instance, the $S$ matrix using (\ref{st}) anyway, one obtains a theory
which is identical to the usual theory at tree level but
not at higher orders. Precisely, one can show (through straightforward
complex integration) that all diagrams containing loops are equal to
zero. This is not in contradiction with the optical theorem because
the $S$ matrix in this context is not unitary but Krein-unitary. The
above property then reflects the fact that the Krein-unitarity is a
different requirement than unitarity. Consequently, the implementation
of unitarity in the Krein space context leads to a modification of
usual interacting terms.

The sums in Eqs.(\ref{kunita1}, \ref{kunita2}) extend over the whole
Fock space whereas only the states which belong to
$\underline{{\cal H}_p}$ \ie, with positive norm, are physical states.
Thus,
only these states
can pretend to be  the so-called ``in'' states of the interacting theory.
In
other words, the space of ``in'' states
cannot be isomorphic to the whole Fock space
$\underline{\cal H}$ but only to its subspace $\underline{{\cal H}_p}$.
Since in addition the space of ``out'' states
is in one-to-one
correspondence with the space of ``in'' states we conclude that a
physically admissible $S$-matrix
cannot connect physical states to unphysical ones. Finally the 
$S$-matrix must satisfy
\begin{equation}\label{condunita}
\left\{
\begin{array}{l}
{\displaystyle
\sum_{\{\alpha_+\}} \left |\bra{\alpha_+} S \ket{\alpha}
\right |^2 \,= \,\sum_{\{\alpha_+\}}
\left |\bra{\alpha} S \ket{\alpha_+}
\right |^2 \,= \,1
}\\
{\displaystyle
\left |\bra{\alpha_-^{(n)}} S \ket{\alpha}
\right |^2\,=\,\left |\bra{\alpha} S \ket{\alpha_-^{(n)}}
\right |^2\,=\,0,
}
\end{array}
\right.
\end{equation}
where $\alpha\in\underline{\cal H}$. 

This can be achieved in the following way.  Let $\Pi_+$ be 
the projection over
$\underline{{\cal H}_p}$: $$
\Pi_+ \,=
\,\sum_{\{\alpha_+\}} \ket{\alpha_+}\bra{\alpha_+}\, .
$$

Indeed Eq.(\ref{quantumfield}) reads as
\begin{equation}\label{quantphimik}
\begin{split}
\varphi(x)\,&= \,\sum_{k} \Bigl\{
\bigl(a_k\,\phi_k(x)\,+\,a^\dag_k\,\phi_k^*(x)\bigr)
\\
& -\bigl(b^\dag_k\,\phi_k(x)\,+\,b_k\,\phi_k^*(x)\bigr)
\Bigr\}\\
& = \varphi_+ -
\varphi_- ,
\end{split}
\end{equation}
and this defines the ($\varphi_+$) and 
($\varphi_-$) parts.  Then one has
\begin{equation}
\Pi_+ \varphi \Pi_+ \ket{\gamma} =
\begin{cases}
\varphi_+\ket{\gamma}&  if \ket{\gamma} \in \underline{{\cal
H}_p}\\
0,& if \ket{\gamma} \in \bigl(\underline{{\cal H}_p}\bigr)^\perp.
\end{cases}
\end{equation}
Consequently, we start with $V(\Pi_+\varphi\Pi_+)$ instead of
$V(\varphi)$ in deriving the $S$-matrix.

By following this quantization procedure for the interaction term, the
k-scalar field theory will lead to the same predictions as the usual
scalar field theory. The only difference lies in the vacuum energy
density: in the k-scalar field theory the free vacuum energy density
is always zero. When interaction is present the vacuum energy density
receives (divergent) contributions of higher order vacuum bubble
graphs.  Note that the so-called radiative corrections are the same in
both theories. In this respect vacuum effects in k-scalar field theory
only involve the interacting vacuum. This may appears as an advantage
if one believes that a truly free theory is a theoretical fiction.

Note that the new interaction term $V(\Pi_+\varphi\Pi_+)$ is in some
sense the quantum version of the restriction $V'$ of $V$ to the
positive energy $\phi$'s. In this sense the Lagrangian density should
be written
\begin{equation*}
{\cal L}=
g^{\mu\nu}\partial_\mu\phi\partial_\nu\phi-m^2\phi-V'(\phi).
\end{equation*}

The Krein quantization of other canonically quantizable theories
may be carried out following the same lines. A prescription for
interacting terms is to replace the various fields $\xi$ by their
restriction $\Pi_+^\xi \xi \Pi_+^\xi$, where $\Pi_+^\xi$ is the
corresponding projector. Such a k-field theory,
will only differ from the usual one by the vanishing of the
free field vacuum energy.

Now, we must underline that, in a curved space, the operator $\Pi_+$
may not exist, and the problem of interacting field on such a space
is, as is well known, much more elaborated. Note that, in particular,
the function $G^{(1)}(x, x') = \bra{0} \{\varphi(x),\varphi(x')\}
\ket{0}$ vanishes in our formalism, and thus cannot have the Hadamard
property. This could be a problem for the construction of perturbative
interacting field theories in curved space \cite{brunefred,
holwal}. Nevertheless, the link between the vacuum and the two-points
function is not exactly the same in the usual quantization scheme and
in the Krein space quantization context: in the usual approach to
choose a vacuum is to choose a physical space of states and a
two-points function, in our approach the vacuum is unique and
consequently do not determines the physical space of states but the
link between this space and the two-points function
remains\cite{grt}. As a consequence a two-points function with
Hadamard property is still present but with another meaning. The 
description of an interacting theory using the Krein-space quantization 
approach remains open. 
  
\subsection*{ACKNOWLEDGMENTS}
The authors thank J-P. Gazeau for valuable discussions.

\subsection*{APPENDIX: THE FOCK SPACE}

Any Krein space $\H$ is equipped with a (non unique) Hilbert space
structure.  Hence, one can define over $\H$ a Fock space through
\cite{mintchev} 
$$\underline{\cal H}=\bigoplus_{n\geq0}S_n({\cal
H}),
$$ 
where $S_n({\cal H})$ is the $n^{\rm th}$ symmetric tensor
product $\cal H$. In a Fock space creation and annihilation operators are
well-defined, not only for the modes but also for any element of
$\H$. Suppose that the scalar product is defined through (\ref{pskg}),
one has:
\begin{multline*}
(a(\phi)\Psi)(x_1,\ldots,x_{n-1}) =  \\
\sqrt{n} \int_{\Sigma}
\phi^{*}(x)
\stackrel{\leftrightarrow}{\partial_{\mu}}\Psi(x,x_1,\ldots,
x_{n-1})d\sigma^\mu(x),
\end{multline*}
 for any $\Psi\in
S_{n}(\uh)$. The creation operator reads
\begin{multline*}
(a^{\dag}(\phi)\Psi)(x_1,\ldots,x_{n+1})= \\
\frac{1}{\sqrt{n+1}}
\sum_{i=1}^{n+1}\phi(x_i)\Psi(x_1,\ldots,\check{x_i},\ldots,x_{n+1}).
\end{multline*}

One  verifies that:
\begin{eqnarray*}
[a(\phi),a(\phi')]&= &0 = [a^{\dag}(\phi),a^{\dag}(\phi')] =  0\\[0pt]
[a(\phi),a^{\dag}(\phi')]&=  &\le\phi,\phi'\re,
\end{eqnarray*}
and
\begin{equation}
\lbl{cov}\uu_ga^{\dag}(\phi)\uu_g^*=a^{\dag}(U_g\phi),\mbox{ et
} \uu_ga(\phi)\uu_g^*=a(U_g\phi),\end{equation}
where $U$ is the natural representation of the group $G$ on 
$\H$ and $\uu$ is the extension of this representation to the Fock space.


\begin{thebibliography}{AAA} \baselineskip=10pt
\bibitem{grt}
J-P. Gazeau,  J. Renaud, M. Takook,
 Class. Quantum Grav. {\bf17}, 1415, (2000).
\bibitem{birreldavis} N.D.~Birrell, P.C.W.~Davies, \emph{Quantum fields in
curved space}. Cambridge University Press, 1982.
\bibitem{wald}  R.M. Wald, \emph{Quantum Fields Theory in
Curved Spacetime and Black Hole Thermodynamics}. The University of 
Chicago Press, 1994.
\bibitem{garhugren} T.~Garidi, E.~Huguet, J.~Renaud,
Phys. Rev.~D {\bf 67}, 124028, (2003), and references herein.
\bibitem {allen1} B.~Allen, Phys. Rev.~D {\bf 32}, 3136 (1985).
\bibitem {chernikov} N.~A.~Chernikov, E.~A.~Tagirov, Ann. Inst. Henri
Poincar\'e,  {\bf 9}, 109 (1968).
\bibitem {mottola} E.~Mottola, Phys. Rev.~D {\bf 31}, 754 (1985).
\bibitem {daniel1} U.~H~Danielsson, JHEP {\bf 0212}, 025  (2002).
\bibitem {daniel2} U.~H~Danielsson, JHEP {\bf 0207}, 040  (2002).
\bibitem {antoniadis} I.~Antoniadis, P.~O.~Mazur, E.~Mottola,~
astro-ph/$9705200$.
\bibitem {mazur} P.~O.~Mazur, E.~Mottola,
Phys. Rev.~D {\bf 64}, 104022 (2001).
\bibitem {bousso} R.~Bousso, A.~Maloney, A.~Strominger, 
Phys. Rev.~D {\bf 65}, 104039 (2002).
\bibitem {tolley} A.~Tolley, N.~Turok, hep-th/0108119.
\bibitem {banksmannelli} T.~Banks, L.~Mannelli
Phys. Rev.~D {\bf 67}, 065009, (2003).
\bibitem {einhornlarsen} M. B.~Einhorn, F.~Larsen, 
Phys. Rev.~D {\bf 67}, 024001, (2003).
\bibitem {goldsteinlowe} K.~Goldstein, D.~A.~Lowe, Nucl. Phys. {\bf B669},
325, (2003). 
\bibitem{higu1}   A. Higuchi,
Class. Quant. Grav. {\bf 8},  1961 (1991).
\bibitem{higu2}   A. Higuchi,
Class. Quant. Grav. {\bf 8},  1983 (1991).
\bibitem{marolf}    D. Marolf, ``Group Averaging and Refined Algebraic Quantization: Where are we now?'' {\it in} Proceedings of the 9th Marcel Grossmann Conference, Rome 2000.
\bibitem{fulling} S. A. Fulling, \emph{Aspects of Quantum Field Theory
in Curved Spacetime}, Cambridge University Press, Cambridge (1989)
\bibitem{isham} C. J. Isham, ``Quantum field theory in curved
spacetimes: A general mathematical framework'' {\it in}
\emph{Differential geometrical methods in mathematical physics II}, Lecture
Notes in Math., vol 676, ed. by K. Bleuler {\it et alii}, Springer,
Berlin (1978).
\bibitem{mintchev}M. Mintchev,  J.
Phys. A, {\bf 13}, 1841, (1990).
\bibitem{bognar} J. Bognar, \emph{Indefinite inner product
spaces}, Springer-Verlag, Berlin (1974).
\bibitem{dbr2} S. De Bi\`evre, J. Renaud, 
Phys. Rev. D  {\bf 57}, 6230, (1998).
\bibitem{dbr3} S. De Bi\`evre, J. Renaud
J. phys. A  {\bf 34},  10901 (2001).
\bibitem{brunefred} R. Brunetti, K. Fredenhagen,
Commun. Math. Phys., 208, (2000).
\bibitem{holwal} R. Hollands, R. Wald, Commun. Math. Phys., 231, (2002).
\end{thebibliography}
\end{document}